\documentclass{article}

\usepackage[final]{neurips_2023}

\usepackage[utf8]{inputenc} 
\usepackage[T1]{fontenc}    
\usepackage{hyperref}       
\usepackage{url}            
\usepackage{booktabs}
\usepackage{amsfonts}       
\usepackage{nicefrac}       
\usepackage{microtype}      
\usepackage{xcolor}         

\usepackage{graphicx}
\usepackage{tabularx} 
\usepackage{ragged2e} 
\usepackage{threeparttable}
\usepackage{amsmath}
\usepackage{threeparttable}
\usepackage{bbm}
\usepackage{enumitem}

\usepackage{array} 
\usepackage{booktabs} 
\usepackage{multirow} 
\usepackage{bbding}
\usepackage{pifont}

\usepackage[square,numbers]{natbib}
\title{Step-Audio-AQAA: a Fully End-to-End Expressive Large Audio Language Model}

\author{%
  Step-Audio Team \\
  StepFun\\
}

\begin{document}

\maketitle

\begin{abstract}
Large Audio-Language Models (LALMs) have significantly advanced intelligent human-computer interaction, yet their reliance on text-based outputs limits their ability to generate natural speech responses directly, hindering seamless audio interactions. To address this, we introduce Step-Audio-AQAA, a fully end-to-end LALM designed for Audio Query-Audio Answer (AQAA) tasks. The model integrates a dual-codebook audio tokenizer for linguistic and semantic feature extraction, a 130-billion-parameter backbone LLM and a neural vocoder for high-fidelity speech synthesis. Our post-training approach employs interleaved token-output of text and audio to enhance semantic coherence and combines Direct Preference Optimization~(DPO) with model merge to improve performance. Evaluations on the StepEval-Audio-360 benchmark demonstrate that Step-Audio-AQAA excels especially in speech control, outperforming the state-of-art LALMs in key areas. This work contributes a promising solution for end-to-end LALMs and highlights the critical role of token-based vocoder in enhancing overall performance for AQAA tasks. The model and demo have been released \footnote[1]{Hugging Face: \url{https://huggingface.co/stepfun-ai/Step-Audio-AQAA}}.
\end{abstract}

\section{Introduction}
\label{introduction}

Large language models (LLMs) have significantly advanced intelligent human-computer interaction, spanning scenarios such as knowledge-based question answering \cite{li2024flexkbqa, claude_3_5, Gemini, grattafiori2024llama}, code assistance~\cite{nam2024using, copilot}, affective companionship \cite{jaech2024openai, hurst2024gpt} and multi-modal interaction~\cite{copet2023simple,wang2024lami,zhuang2025vistorybench,hu2024drivingworld}. The integration of auxiliary techniques — including reinforcement learning (RL) \cite{kirk2023understanding, wang2024comprehensive}, tool calling \cite{kim2024llm, zhuang2023toolqa}, and deep search \cite{zou2021pre, xiong2024search} — has further enhanced the factual accuracy and timeliness of LLMs, sparking a wave of research innovation.

Nevertheless, human communication and environmental perception extend beyond textual modalities to encompass speech and audio signals. Unlike text, speech inherently encodes rich paralinguistic cues (e.g., timbre, emotional prosody, intonation, and stress patterns) \cite{nematullayevna2024role, schuller2013paralinguistics}, while non-speech audio provides contextual information deeply intertwined with real-world scenarios \cite{gong2023joint}. Consequently, large audio-language models (LALMs), which refers to LLMs capable of generating intelligent verbal responses based on the input speech or audio \cite{cheng2024videollama, chu2024qwen2, kong2024audio, yang2024uniaudio}, have emerged as a critical milestone toward achieving artificial general intelligence.
And researchers have proposed numerous LALMs that exhibit impressive performance across diverse dimensions \cite{cui2024recent, ji2024wavchat, peng2024survey}, including speech intelligence, audio and music understanding and generation, multilingual capability and even multi-modal capability. 

The initial research on LALMs focused on converting speech modalities into text and establishing functional connections with LLMs. For example, HuggingGPT \cite{shen2023hugginggptsolvingaitasks} decomposed human instructions using LLMs and invoked Huggingface models to perform tasks like automatic speech recognition (ASR), text to speech (TTS), and audio inpainting. Similarly, AudioGPT \cite{huang2024audiogpt} integrated diverse audio foundation models to handle complex audio data and bridged LLMs with ASR/TTS interfaces for speech interactions. However, these approaches relied on cascaded sub-modules with limited functionality and were prone to error accumulation \cite{peng2024survey}.

Later research advanced LALMs by incorporating discrete audio tokens \cite{zeghidour2021soundstream, wu2024codec} or continuous audio features, significantly improving performance in spoken language understanding tasks. A series of VALL-E models \cite{han2024vall, chen2024vall, wang2023neural} and SpeechGPT \cite{zhang2023speechgpt} demonstrated deeper integration of speech and LLMs, enabling both audio processing and natural language interaction. Google’s AudioPaLM \cite{rubenstein2023audiopalm} further extended these capabilities into multi-modal processing. Additionally, broader data annotation and task definitions enhanced LALMs' open-ended and close-ended abilities. For instance, Pengi \cite{deshmukh2023pengi} framed all audio tasks as text-generation problems and benchmarked its performance on 21 downstream tasks, including open-ended tasks like Audio Captioning and AQTA. SALMONN \cite{tang2023salmonn} showcased emergent abilities not explicitly trained for, such as speech translation into untrained languages, audio-based storytelling, and co-reasoning with speech and audio. Similar efforts include Qwen2-Audio \cite{chu2024qwen2}, Qwen2.5-Omni \cite{xu2025qwen2}, GLM-4-Voice \cite{zeng2024glm}, Step-Audio \cite{huang2025step} and Kimi-audio \cite{ding2025kimi}.
Despite these advancements, most of these models output results in text tokens, failing to achieve end-to-end speech understanding and generation.

Motivated by these limitations and the growing prominence of RL in audio generation \cite{gao2025emo, tian2025preference, zhang2024speechalign}, this study introduces Step-Audio-AQAA, where AQAA stands for Audio Query-Audio Answer tasks. Step-Audio-AQAA is a fully end-to-end LALM specifically designed to handle audio queries and comprehension while generating natural, accurate, and low-latency speech responses. The main contributions of this work are summarized as follows:

\begin{itemize}

    \item \textbf{Fully end-to-end speech large model}: Unlike the cascaded approach, our model, Step-Audio-AQAA, directly generates the target output (text/speech) from raw audio input without the need for ASR or TTS. This "pure" end-to-end design not only significantly simplifies system complexity and eliminates cascaded errors, but also demonstrates substantial performance improvements through joint optimization on large-scale speech-text pairing data.


    \item \textbf{Fine-grained voice control ability}: Through carefully designed training strategies and data organization methods, we have achieved fine-grained voice control capabilities in Step-Audio-AQAA, enabling sentence-level modifications such as emotional tone and speech rate. Such capabilities were not attainable with our previous AQTA+TTS paradigm \cite{stepfun_2025}.

\end{itemize}


\section{Architecture}
\label{method}

\begin{figure}[t]
    \centering
    \includegraphics[width=0.8\columnwidth]{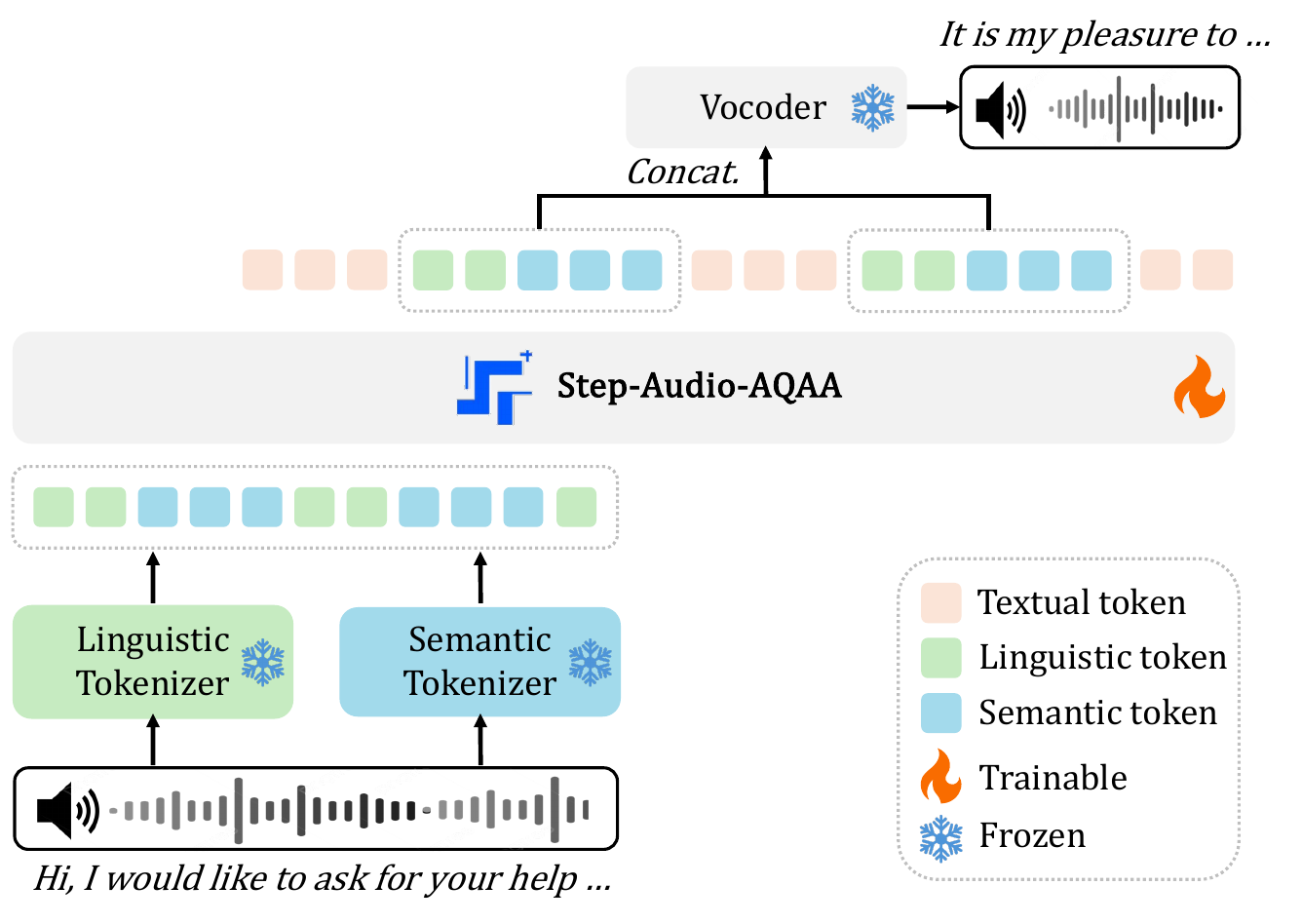}
    \caption{
      Model architecture of Step-Audio-AQAA. The backbone of Step-Audio-AQAA is a pre-trained 130-billion-parameter multi-modal LLM, Step-Omni \cite{huang2025step}, which is further post-trained through SFT and DPO in this study, ultimately evolving into Step-Audio-AQAA system. The audio query is synchronously discretized into linguistic tokens and semantic tokens, which are then merged into an input sequence with a 10:15 interleaving ratio. The output sequence consists of textual tokens and audio tokens, with these tri-codebook tokens interleaved in a 10:6:9 ratio. The vocoder is a flow-matching model that shares a similar architecture with CosyVoice \cite{du2024cosyvoice}, but it is uniquely conditioned solely on the audio tokens.}
    \vspace{-6mm}
    \label{fig:framework}
    
\end{figure}

Step-Audio-AQAA adopts an end-to-end paradigm for audio-language modeling, comprising three core modules: a dual-codebook audio tokenizer, a backbone LLM, and a neural vocoder, as illustrated in Figure \ref{fig:framework}. The system processes audio-modal queries through the following pipeline: (1) Firstly, the dual-codebook audio tokenizer converts the input audio into a hybrid sequence of linguistic tokens and semantic tokens. For brevity, they will be referred as audio tokens. (2) Then, the core LLM, post-trained through SFT, DPO and model merge, generates an output sequence interleaving text tokens and audio tokens. (3) Finally, the vocoder module reconstructs high-fidelity speech waveforms from the generated audio tokens as responses to the input queries. This architecture enables seamless interaction, where audio inputs are transformed into structured token representations, processed by the LLM to produce contextually relevant outputs, and finally rendered as natural speech responses through waveform synthesis.

\subsection{Dual-Codebook Audio Tokenizers}

Step-Audio-AQAA utilized two different tokenizers — linguistic and semantic — to enhance the representation of speech features. The linguistic tokenizer was employed to extract structured, high-level representations, such as phonemic and linguistic attributes, while the semantic tokenizer was intended to encode coarse-grained acoustic characteristics. The reason to use the dual-codebook audio tokenizers was that the linguistic tokens and semantic tokens were mutually referenced, and we observed that when using dual-codebook training, the next token prediction perplexity for both semantic tokens and linguistic tokens decreased compared to using a single codebook in \cite{huang2025step}.

Specifically, the linguistic tokenizater used the output from the Paraformer encoder \cite{gao2022paraformer}, quantized into discrete tokens at a rate of 16.7 Hz with a codebook size of 1,024, while the semantic tokenizater was refer to CosyVoice 1.0 \cite{du2024cosyvoice}, designed to efficiently encode features critical for speech synthesis, operating at 25 Hz with a larger codebook size of 4,096 to capture finer acoustic details. Since the sampling rates of the two types of tokens were approximately in a 2:3 ratio, we adopted a 2:3 interleaving ratio to ensure temporal alignment of tokens, thereby forming the final input sequence for the LLM, as shown in the output side of Figure \ref{fig:framework}.

\subsection{Backbone LLM} 
In order to enhance the ability of speech understanding and the semantic consistency of generation in a cost-effective manner, we chose the backbone LLM as a pre-trained 130-billion-parameter multi-modal LLM, Step-Omni \cite{huang2025step}, whose pre-training data spans three modalities: text, speech, and image. The embedding layer's vocabulary was extended by incorporating 5,120 audio tokens into the pre-trained text vocabulary, followed by the integration of a pre-trained image encoder. Noted that this study only utilized the text and speech capabilities of Step-Omni in the post-training stage and inference stage.

Step-Omni employed a decoder-only architecture. In this architecture, the dual-codebook audio tokens were first embedded using the merge vocabulary, followed by multiple Transformer blocks. Each Transformer block consisted of an input RMSNorm layer \cite{zhang2019root}, a grouped query attention module, a post-attention RMSNorm layer, and a feed-forward layer. Finally, the model concluded with a final RMSNorm layer and a linear language modeling head.

The multi-modal pre-training process of Step-Omni will be elaborated in detail in the Subsection \ref{llm_pretraing}. Subsequently, the pre-trained Step-Omni was further adapted for the AQAA task through a post-training stage, including SFT, DPO, and model weight merging, ultimately evolving into the proposed Step-Audio-AQAA model. 
The post-trained LLM produced contextually relevant outputs composed of interleaved textual and audio tokens at a 10:15 ratio\footnote{Use a TTS model to convert text into speech tokens. Interleave text tokens and speech tokens in a 10:15 ratio. If the number of text tokens is insufficient, fill the remaining positions with speech tokens.}. Notably, during the post-training phase of DPO, we intentionally retained textual tokens as part of the output to leverage their auxiliary role in facilitating objective function convergence (as detailed in Section \ref{dpo}). This tri-Codebook post-training enhances semantic consistency in the generated audio tokens by leveraging multi-modal alignment cues from textual representations.

\subsection{Neural Vocoder} 

The generated audio tokens were synthesized into natural, high-quality speech via a vocoder and returned to the user. The vocoder in this study draws inspiration from the open-source optimal-transport conditional flow matching model introduced in CosyVoice 1.0 \cite{du2024cosyvoice}, which employs a U-Net architecture with basic modules integrating ResNet-1D~\cite{he2016deep} layers and Transformer blocks for efficient feature extraction and temporal modeling.

\section{Training and Dataset}

\begin{figure}[t]
    \centering
    \includegraphics[width=\columnwidth]{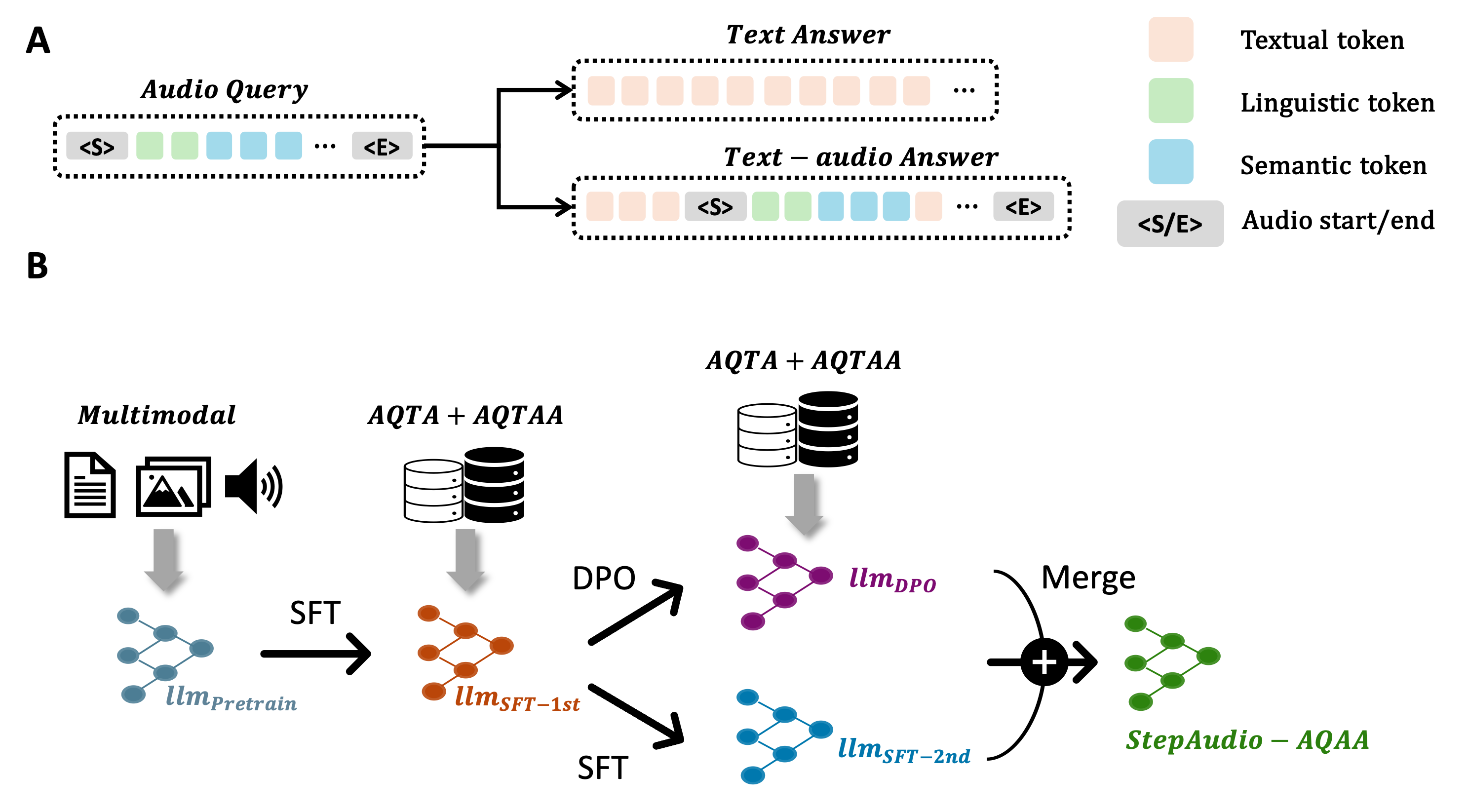}
    \caption{Illustration of (A) tokenized AQTA data pairs and tokenized AQTAA data pairs utilized in the superivised fine-turning stage, and (B) mutli-stage model training process. Consistent with Figure \ref{fig:framework}, the audio tokens are composed by linguistic tokens and semantic tokens, with a 2:3 interleaving ratio, while the tokens of text-audio answer are interleaved in a 3:2:3 ratio. SFT: Superivised Fine-turning; DPO: Direct Preference Optimization; AQTA: Audio Query-Text Answer dataset; AQTAA: Audio Query-Text Answer-Audio Answer dataset.}
    \label{fig:train_data}
    
\end{figure}

\subsection{LLM Pre-Training}
\label{llm_pretraing}

It is consistent with the pre-training method in \cite{stepfun_2025}. The pre-training dataset for Step-Omni encompasses three modalities: audio, text, and images. Specifically, the text data, along with image-text paired and alternating data, is sourced from web pages, books, and proprietary resources,  amounting to 800 billion tokens separately. While the audio modality consists of several types of data, including audio continuation sequences, TTS synthesized speech, ASR data, and audio-text alternating data.

The multi-modal pre-training process is divided into three distinct stages. Firstly, the training data is utilized in a ratio of 2:1:1 for audio, text, and image modalities, respectively. During this phase, model parameter updates are primarily concentrated on the embedding layers and LM head associated with the audio modality. In the second stage, audio-text interleaved data is incorporated to further enhance the audio performance. Finally, in the third stage, ASR and TTS data are introduced for additional pre-training. Notably, this staged approach ensures that the model progressively refines its multi-modal capabilities while maintaining the textual ability. 

\subsection{Supervised Fine-Tuning}

After completing the pre-training phase, we conducted two stage supervised fine-tuning in both Audio Query-Text Answer (AQTA) and Audio Query-Text Answer-Audio Answer (AQTAA) formats. The AQTA data is proprietary, while the AQTAA dataset is generated based on the AQTA data, during which the Step-Audio-TTS-3B model \cite{huang2025step} converted text-based answers into high-quality audio responses. And the token organization during training for the two types of data pair is illustrated in the Figure \ref{fig:train_data}. 

In the first stage of SFT, the full parameters of pre-trained LLM were updated on the combined AQTA and AQTAA datasets for one epoch.
This was aimed at enhancing the model's semantic consistency in question-answering scenarios and aligning its input-output structure with the end-to-end audio interaction paradigm.
In the subsequent stage, to further stabilize the output format of the LLM to a text-audio interleaved structure and enhance certain abilities, such as singing, we selected some high-quality AQTAA data and trained it for a certain number of steps. The objective function in the two stage is the  cross-entropy (CE) loss, which computes loss only for the tokens in the response part:
\begin{align}
\mathcal{L}_{\text{CE}}(\theta) = -\frac{1}{T} \sum_{t=1}^{T} \log P_{\theta}(y_t | x, y_{<t}),
\end{align}

where $\theta$ represents the parameters of the LLM, $x$ denotes the input prompt, and $y$ corresponds to the target response sequence.

\subsection{Direct Preference Optimization}
\label{dpo}
To further align the model's outputs with human preferences and enhance its generalization capability, we explored DPO \cite{rafailov2023direct}. 

\paragraph{Audio-token Masked Direct Preference Optimization}

In our study, the generation of text-audio interleaved responses operates at the token level. To align LLMs with human preferences through token-level policy optimization. We discovered that applying DPO optimization to all tokens resulted in sub-optimal effects, specifically manifesting as some text and audio misalignment. We suspect that DPO partially compromised the ability to generate voice tokens, so during subsequent DPO processes, we blocked the loss of audio tokens. 
The masked-DPO loss function is formulated as follows:
\begin{align}
L_{mDPO} &= -\mathbb{E}_{(s_0, \tau^w, \tau^l) \sim D} \log \sigma \nonumber \\
&\quad  \left[ \sum_{t=0}^{T_w - 1} \beta \mathbbm{I}(a_t^w \notin A)  \log \frac{\pi_\theta(a_t^w | s_t^w)}{\pi_{ref}(a_t^w | s_t^w)} \right. \nonumber \\
&\quad \left. - \sum_{t=0}^{T_l - 1} \beta \mathbbm{I}(a_t^l \notin A) \log \frac{\pi_\theta(a_t^l | s_t^l)}{\pi_{ref}(a_t^l | s_t^l)} \right],
\end{align}
where:
\begin{itemize}

    \item $(s_t^w, a_t^w )$ and $(s_t^l, a_t^l)$ denote state-action pairs at time $t$ in the preferred and dis-preferred trajectories, respectively;

    \item $A$ denotes set of audio tokens and $T$ is the trajectory length;
    
    \item $\beta$ controls the deviation from the base reference policy $\pi_{ref}$;

\end{itemize}

We started DPO from the first-stage SFT model because the second-stage SFT model specially enhanced certain abilities, thereby causing damage to some other abilities.

\subsection{Weight Merging}

Given the distinct optimization objectives of the SFT-first-stage model, SFT-second-stage model, DPO-fine-tuned model, finally, we integrate the three backbone LLM variants through weighted averaging of their parameter matrices \cite{matena2022merging,wortsman2022model, cao2022ml4co}. This ensemble strategy aims to enhance answer accuracy and semantic consistency by leveraging complementary strengths across models. The resulting merged model serves as the final backbone LLM for Step-Audio-AQAA.

As shown in Equation \ref{eq:weight merging}, weight merging is achieved by performing weighted averaging of the parameter matrices $W$ at corresponding positions across individual models:
\begin{align}
W_{Step-Audio\mbox{-}AQAA} = (5*W_{SFT-1st} + 5*W_{SFT-2ed} + 1*W_{DPO})/11.
\label{eq:weight merging}
\end{align}



\section{Evaluation Setup}
\label{exp}

\subsection{Benchmark}
\label{bmk}

StepEval-Audio-360 \cite{stepfun_2025} is a comprehensive benchmark dataset designed to evaluate the capabilities of LALMs in human-AI audio interaction. Sourced from professional human annotators, this dataset spans a wide range of skills, including singing, creativity, role-playing, logical reasoning, voice instruction following, voice understanding, gaming, speech emotion control, and language ability. 
The dataset features human voice recordings in multiple languages and dialects, such as Chinese (including Szechuan and Cantonese dialects), English, and Japanese, ensuring diversity in linguistic and acoustic contexts. StepEval-Audio-360 has been released at \url{https://huggingface.co/datasets/stepfun-ai/StepEval-Audio-360}.

\subsection{Baselines and Metrics}

Kimi-Audio \cite{ding2025kimi} and Qwen-Omni \cite{xu2025qwen2} are end-to-end LALMs capable of directly understanding and generating both Chinese and English speech. They support real-time voice interactions and can adapt speech attributes such as emotion, tone, speed, and dialect based on user instructions. Therefore, they can represent the state-of-the-art performance of LALMs in AQAA tasks.

To assess model performance, expert evaluators rated end-to-end dialog sessions using a 1-5 Mean Opinion Score (MOS) scale for naturalness and task completion. Comprehensive human evaluations were conducted to compare Step-Audio-AQAA with Kimi-Audio and Qwen-Omni across the nine critical dimensions of StepEval-Audio-360 outlined above. This rigorous evaluation highlights the strengths and limitations of each model in delivering high-quality audio interactions.

\begin{figure}[t]
    \centering
    \includegraphics[width=1\columnwidth]{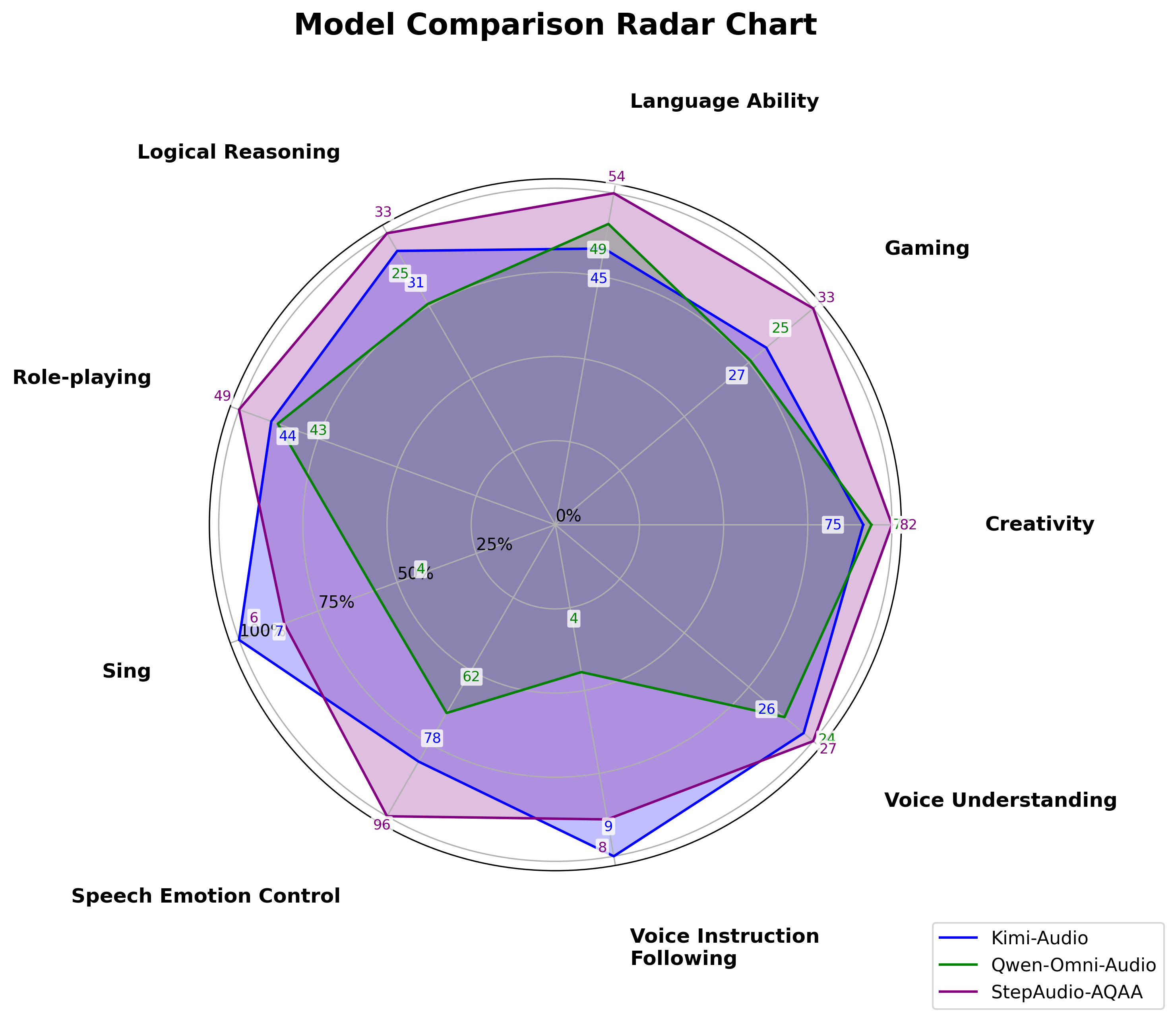}
    \caption{Human evaluation of the end-to-end speech interactions on StepEval-Audio-360 benchmark. The benchmark can be categorized into nine categories, and the radar chart illustrates the total MOS scores of the three LALMs across each category, respectively.
      }
    
    \label{fig:radar}
\end{figure}


\section{Results and Discussions}

\subsection{MOS Scores on StepEval-Audio-360}
The evaluation results on the StepEval-Audio-360 benchmark reveal distinct performance profiles among the three models, as illustrated in Figure \ref{fig:radar}.

Step-Audio-AQAA demonstrates a leading edge across multiple key dimensions. Notably in Speech Emotion Control, Step-Audio-AQAA showed its superior ability in expressing and recognizing vocal emotions. In Creativity, Language Ability, Gaming and Role-playing, Step-Audio-AQAA also achieved the highest scores, indicating its comprehensive strength in understanding complex instructions, generating diverse content, and engaging in fluent interactions. For Logical Reasoning and Voice Understanding, Step-Audio-AQAA also led, although its advantage in the latter was relatively marginal.

Step-Audio-AQAA has certain disadvantages in the two dimensions of Singing and Voice Instruction Following. This is because adding excessive sing data to enable the model to learn singing will seriously damage other capabilities; meanwhile, the lack of data similar to Voice Instruction Following also leads to the model's weak performance in this capability. We will leave these optimizations for the future.




\subsection{Ablation Study}
We further explored the influence of two training settings: text-audio token mixing proportions and text-audio interleaving methods. In order to save manpower and maintain objectivity, we adopt LLM as the judger for automatic evaluation of these ablation studies. Specifically, we use GPT-4o as the judge to score the model's responses in three dimensions: chat, relevance and factuality.

\paragraph{Different Text-Audio Token Mixing Proportions} we aim to explore the impact of varying the mixing proportions of audio tokens on the performance of our model. We set up multiple experimental groups, each with a distinct ratio of audio tokens from same sources: 
(1) ratio\_10\_15: text-audio token ratio is 10:15; (2) ratio\_6\_50: text-audio token ratio is 6:50; (3) ratio\_6\_50: text-audio token ratio is 3:5; (4) text\_cot: output all text first, then output audio tokens; (5) audio\_only: output audio tokens only.
The results are listed in Table \ref{tab:adjusted_relevance}. As is evident from the table, when the token information of the generated text adequately encompasses the subsequently produced speech tokens, there is a notable enhancement in quality.

\begin{table}[ht]
\centering
\caption{Experimental results with different Interleaving ratios} 
\label{tab:adjusted_relevance} 
\begin{tabular}{lrrr}
\toprule
\textbf{Model} & \textbf{Chat↑} & \textbf{Relevance↑} & \textbf{Factuality↑} \\
\midrule
audio\_only & 1.7158 & 0.0526 & 0.0316 \\
ratio\_6\_50 & 1.2000 & 0.0421 & 0.0526 \\
ratio\_3\_5 & 1.0316 & 0.0000 & 0.0000 \\
text\_cot & 4.0105 & 0.5895 & 0.5789 \\
ratio\_10\_15 & \textbf{4.0316} & \textbf{0.6526} & \textbf{0.6737} \\
\bottomrule
\end{tabular}
\end{table}


\paragraph*{Text-Audio Interleaving Methods}
If the speech output by a model in one turn has a single speech state (such as emotion and speech rate), it is called single-label; if there are multiple speech states, it is called multi-label. For single-label data or unlabeled speech data, we use a TTS model to convert text into speech, where the speech is transformed into a form of tokens similar to ``\textless{}audio\_start\textgreater{}\ldots\textless{}audio\_end\textgreater{}''. To enable the model to have the ability to switch speech states within a turn, we first synthesize single-label audio and then splice it in a certain way to obtain multi-label data. Specifically, when processing multi-label speech data, we considered the following three splicing methods:

\begin{enumerate}
    \item \textbf{concatenation with marker removal}: Before splicing, remove all special audio markers \textless{}audio\_start\textgreater{} and \textless{}audio\_end\textgreater{}, connect the audio tokens in order, then add \textless{}audio\_start\textgreater{} and \textless{}audio\_end\textgreater{} at the beginning and end, and finally interleave them with text tokens at a ratio of 10:15.
    \item \textbf{pre-interleaved concatenation}: First interleave the audio tokens of each label with text tokens at a ratio of 10:15, and then concatenate them in order.
    \item \textbf{marker-preserving concatenation}: Without removing special markers before concatenating, concatenate them in order and then interleave them with text tokens at a ratio of 10:15.
\end{enumerate}

\begin{table}[htbp]
    \centering
    \caption{Results of Different Audio-Text Interleaving Methods}
    \label{tab:interleaving_results} 
    \begin{tabular}{lccc} 
        \toprule 
        \textbf{Method} & \textbf{Chat$\uparrow$} & \textbf{Relevance$\uparrow$} & \textbf{Factuality$\uparrow$} \\
        \midrule 
        pre-interleaved concatenation & 3.8211 & 0.5368 & \textbf{0.5895} \\
        concatenation with marker removal & 4.0842 & 0.5579 & 0.5684 \\
        marker-preserving concatenation & \textbf{4.2211} & \textbf{0.5684} & 0.5684 \\
        \bottomrule 
    \end{tabular}
\end{table}

In the initial concept, we hope to adopt a curriculum-learning approach, where we first learn single-label data in the first stage and then multi-label data in the second stage\footnote{In our final training plan, we also incorporated multi-label data into the first SFT stage.}. Therefore, based on the Stage-1 SFT model, we performed training by incorporating the above-mentioned data into the Stage-2 SFT data, and the results that shown in the Table \ref{tab:interleaving_results} indicate that \textbf{marker-preserving concatenation} is the most effective method. 

In addition, we also found that after training with \textbf{concatenation with marker removal} and \textbf{pre-interleaved concatenation}, the model hardly generates multi-label speech. This is because the single-label data all adopt a 10:15 mixing method, and the \textbf{pre-interleaved concatenation} approach disrupts this consistency, increasing the difficulty for the model to learn. Additionally, since the model is learned to maintaining a single speech state within the "\textless{}audio\_start\textgreater{}...\textless{}audio\_end\textgreater{}" markers, the \textbf{concatenation with marker removal} method also breaks this consistency. \textbf{Marker-preserving concatenation}, by contrast, avoids these issues, making it the best choice overall.

\section{Conclusion}

In this paper, we tackled the challenges faced by current LALMs in directly generating natural speech responses for AQAA tasks. We presented Step-Audio-AQAA, a pioneering end-to-end LALM enabling seamless and natural audio interactions. Our approach incorporated innovative post-training techniques, including tri-codebook optimization and advanced preference alignment methods like DPO, two-stage SFT and model merge, which significantly improved the model’s semantic coherence and alignment with human preferences.

Evaluations conducted on the StepEval-Audio-360 benchmark revealed that Step-Audio-AQAA surpasses existing models in critical aspects, including speech emothion control, role-playing, creativity, and voice understanding. This work marks a significant advancement in the field of end-to-end speech interaction systems and highlights the promise of text-audio interleaved output pattern and RL on the AQAA tasks.  

\section{Future Direction}

Despite significant progress in audio generation and speech modeling, several important challenges remain unresolved. First, it is still unclear whether meaningful audio tokens can be generated directly without reliance on text token guidance, which may limit the flexibility and applicability of current models in fully unsupervised or non-linguistic audio generation scenarios. Second, while discrete audio tokens have become a dominant paradigm in neural audio modeling, it remains an open question whether they represent the optimal representation for capturing the continuous and nuanced characteristics of natural audio. Third, generating high-quality singing with stable pitch control and rich melodic variation continues to pose technical challenges, particularly in maintaining coherence across long-range musical structures. 

Finally, an intriguing direction for future work lies in exploring whether large speech models can also benefit from advanced inference paradigms such as o1-style~\cite{jaech2024openai} reasoning, potentially enabling more intelligent and context-aware speech synthesis. Addressing these questions will not only advance the theoretical understanding of audio modeling but also enhance the practical capabilities of next-generation speech and audio generation systems.

{
\small
\bibliography{ref}

\begin{thebibliography}{55}
\providecommand{\natexlab}[1]{#1}
\providecommand{\url}[1]{\texttt{#1}}
\expandafter\ifx\csname urlstyle\endcsname\relax
  \providecommand{\doi}[1]{doi: #1}\else
  \providecommand{\doi}{doi: \begingroup \urlstyle{rm}\Url}\fi

\bibitem[Anthropic(2024)]{claude_3_5}
Anthropic.
\newblock Claude 3.5 sonnet.
\newblock 2024.
\newblock URL \url{https://www.anthropic.com/news/claude-3-5-sonnet}.

\bibitem[Cao et~al.(2022)Cao, Xu, Huang, and Zhou]{cao2022ml4co}
Zixuan Cao, Yang Xu, Zhewei Huang, and Shuchang Zhou.
\newblock Ml4co-kida: Knowledge inheritance in dataset aggregation.
\newblock \emph{arXiv preprint arXiv:2201.10328}, 2022.

\bibitem[Chen et~al.(2024)Chen, Liu, Zhou, Liu, Tan, Li, Zhao, Qian, and Wei]{chen2024vall}
Sanyuan Chen, Shujie Liu, Long Zhou, Yanqing Liu, Xu~Tan, Jinyu Li, Sheng Zhao, Yao Qian, and Furu Wei.
\newblock Vall-e 2: Neural codec language models are human parity zero-shot text to speech synthesizers.
\newblock \emph{arXiv preprint arXiv:2406.05370}, 2024.

\bibitem[Cheng et~al.(2024)Cheng, Leng, Zhang, Xin, Li, Chen, Zhu, Zhang, Luo, Zhao, et~al.]{cheng2024videollama}
Zesen Cheng, Sicong Leng, Hang Zhang, Yifei Xin, Xin Li, Guanzheng Chen, Yongxin Zhu, Wenqi Zhang, Ziyang Luo, Deli Zhao, et~al.
\newblock Videollama 2: Advancing spatial-temporal modeling and audio understanding in video-llms.
\newblock \emph{arXiv preprint arXiv:2406.07476}, 2024.

\bibitem[Chu et~al.(2024)Chu, Xu, Yang, Wei, Wei, Guo, Leng, Lv, He, Lin, et~al.]{chu2024qwen2}
Yunfei Chu, Jin Xu, Qian Yang, Haojie Wei, Xipin Wei, Zhifang Guo, Yichong Leng, Yuanjun Lv, Jinzheng He, Junyang Lin, et~al.
\newblock Qwen2-audio technical report.
\newblock \emph{arXiv preprint arXiv:2407.10759}, 2024.

\bibitem[Copet et~al.(2023)Copet, Kreuk, Gat, Remez, Kant, Synnaeve, Adi, and Défossez]{copet2023simple}
Jade Copet, Felix Kreuk, Itai Gat, Tal Remez, David Kant, Gabriel Synnaeve, Yossi Adi, and Alexandre Défossez.
\newblock Simple and controllable music generation.
\newblock In \emph{Thirty-seventh Conference on Neural Information Processing Systems}, 2023.

\bibitem[Cui et~al.(2024)Cui, Yu, Jiao, Meng, Zhang, Wang, Guo, and King]{cui2024recent}
Wenqian Cui, Dianzhi Yu, Xiaoqi Jiao, Ziqiao Meng, Guangyan Zhang, Qichao Wang, Yiwen Guo, and Irwin King.
\newblock Recent advances in speech language models: A survey.
\newblock \emph{arXiv preprint arXiv:2410.03751}, 2024.

\bibitem[Deshmukh et~al.(2023)Deshmukh, Elizalde, Singh, and Wang]{deshmukh2023pengi}
Soham Deshmukh, Benjamin Elizalde, Rita Singh, and Huaming Wang.
\newblock Pengi: An audio language model for audio tasks.
\newblock \emph{Advances in Neural Information Processing Systems}, 36:\penalty0 18090--18108, 2023.

\bibitem[Ding et~al.(2025)Ding, Ju, Leng, Liu, Liu, Shang, Shen, Song, Tan, Tang, et~al.]{ding2025kimi}
Ding Ding, Zeqian Ju, Yichong Leng, Songxiang Liu, Tong Liu, Zeyu Shang, Kai Shen, Wei Song, Xu~Tan, Heyi Tang, et~al.
\newblock Kimi-audio technical report.
\newblock \emph{arXiv preprint arXiv:2504.18425}, 2025.

\bibitem[Du et~al.(2024)Du, Chen, Zhang, Hu, Lu, Yang, Hu, Zheng, Gu, Ma, et~al.]{du2024cosyvoice}
Zhihao Du, Qian Chen, Shiliang Zhang, Kai Hu, Heng Lu, Yexin Yang, Hangrui Hu, Siqi Zheng, Yue Gu, Ziyang Ma, et~al.
\newblock Cosyvoice: A scalable multilingual zero-shot text-to-speech synthesizer based on supervised semantic tokens.
\newblock \emph{arXiv preprint arXiv:2407.05407}, 2024.

\bibitem[Gao et~al.(2025)Gao, Zhang, Chen, Zhang, and Chen]{gao2025emo}
Xiaoxue Gao, Chen Zhang, Yiming Chen, Huayun Zhang, and Nancy~F Chen.
\newblock Emo-dpo: Controllable emotional speech synthesis through direct preference optimization.
\newblock In \emph{ICASSP 2025-2025 IEEE International Conference on Acoustics, Speech and Signal Processing (ICASSP)}, pages 1--5. IEEE, 2025.

\bibitem[Gao et~al.(2022)Gao, Zhang, McLoughlin, and Yan]{gao2022paraformer}
Zhifu Gao, Shiliang Zhang, Ian McLoughlin, and Zhijie Yan.
\newblock Paraformer: Fast and accurate parallel transformer for non-autoregressive end-to-end speech recognition.
\newblock \emph{arXiv preprint arXiv:2206.08317}, 2022.

\bibitem[Gong et~al.(2023)Gong, Liu, Luo, Karlinsky, and Glass]{gong2023joint}
Yuan Gong, Alexander~H Liu, Hongyin Luo, Leonid Karlinsky, and James Glass.
\newblock Joint audio and speech understanding.
\newblock In \emph{2023 IEEE Automatic Speech Recognition and Understanding Workshop (ASRU)}, pages 1--8. IEEE, 2023.

\bibitem[Google(2025)]{Gemini}
Google.
\newblock Gemini 2.0 pro.
\newblock 2025.
\newblock URL \url{https://deepmind.google/technologies/gemini/pro/}.

\bibitem[Grattafiori et~al.(2024)Grattafiori, Dubey, Jauhri, Pandey, Kadian, Al-Dahle, Letman, Mathur, Schelten, Vaughan, et~al.]{grattafiori2024llama}
Aaron Grattafiori, Abhimanyu Dubey, Abhinav Jauhri, Abhinav Pandey, Abhishek Kadian, Ahmad Al-Dahle, Aiesha Letman, Akhil Mathur, Alan Schelten, Alex Vaughan, et~al.
\newblock The llama 3 herd of models.
\newblock \emph{arXiv preprint arXiv:2407.21783}, 2024.

\bibitem[Han et~al.(2024)Han, Zhou, Liu, Chen, Meng, Qian, Liu, Zhao, Li, and Wei]{han2024vall}
Bing Han, Long Zhou, Shujie Liu, Sanyuan Chen, Lingwei Meng, Yanming Qian, Yanqing Liu, Sheng Zhao, Jinyu Li, and Furu Wei.
\newblock Vall-e r: Robust and efficient zero-shot text-to-speech synthesis via monotonic alignment.
\newblock \emph{arXiv preprint arXiv:2406.07855}, 2024.

\bibitem[He et~al.(2016)He, Zhang, Ren, and Sun]{he2016deep}
Kaiming He, Xiangyu Zhang, Shaoqing Ren, and Jian Sun.
\newblock Deep residual learning for image recognition.
\newblock In \emph{Proceedings of the IEEE conference on computer vision and pattern recognition}, pages 770--778, 2016.

\bibitem[Hu et~al.(2024)Hu, Yin, Jia, Deng, Guo, Zhang, Long, and Tan]{hu2024drivingworld}
Xiaotao Hu, Wei Yin, Mingkai Jia, Junyuan Deng, Xiaoyang Guo, Qian Zhang, Xiaoxiao Long, and Ping Tan.
\newblock Drivingworld: Constructingworld model for autonomous driving via video gpt.
\newblock \emph{arXiv preprint arXiv:2412.19505}, 2024.

\bibitem[Huang et~al.(2025)Huang, Wu, Wang, Yan, Hu, Feng, Tian, Shen, Li, Chen, et~al.]{huang2025step}
Ailin Huang, Boyong Wu, Bruce Wang, Chao Yan, Chen Hu, Chengli Feng, Fei Tian, Feiyu Shen, Jingbei Li, Mingrui Chen, et~al.
\newblock Step-audio: Unified understanding and generation in intelligent speech interaction.
\newblock \emph{arXiv preprint arXiv:2502.11946}, 2025.

\bibitem[Huang et~al.(2024)Huang, Li, Yang, Shi, Chang, Ye, Wu, Hong, Huang, Liu, et~al.]{huang2024audiogpt}
Rongjie Huang, Mingze Li, Dongchao Yang, Jiatong Shi, Xuankai Chang, Zhenhui Ye, Yuning Wu, Zhiqing Hong, Jiawei Huang, Jinglin Liu, et~al.
\newblock Audiogpt: Understanding and generating speech, music, sound, and talking head.
\newblock In \emph{Proceedings of the AAAI Conference on Artificial Intelligence}, volume~38, pages 23802--23804, 2024.

\bibitem[Hurst et~al.(2024)Hurst, Lerer, Goucher, Perelman, Ramesh, Clark, Ostrow, Welihinda, Hayes, Radford, et~al.]{hurst2024gpt}
Aaron Hurst, Adam Lerer, Adam~P Goucher, Adam Perelman, Aditya Ramesh, Aidan Clark, AJ~Ostrow, Akila Welihinda, Alan Hayes, Alec Radford, et~al.
\newblock Gpt-4o system card.
\newblock \emph{arXiv preprint arXiv:2410.21276}, 2024.

\bibitem[Jaech et~al.(2024)Jaech, Kalai, Lerer, Richardson, El-Kishky, Low, Helyar, Madry, Beutel, Carney, et~al.]{jaech2024openai}
Aaron Jaech, Adam Kalai, Adam Lerer, Adam Richardson, Ahmed El-Kishky, Aiden Low, Alec Helyar, Aleksander Madry, Alex Beutel, Alex Carney, et~al.
\newblock Openai o1 system card.
\newblock \emph{arXiv preprint arXiv:2412.16720}, 2024.

\bibitem[Ji et~al.(2024)Ji, Chen, Fang, Zuo, Lu, Wang, Jiang, Zhou, Liu, Cheng, et~al.]{ji2024wavchat}
Shengpeng Ji, Yifu Chen, Minghui Fang, Jialong Zuo, Jingyu Lu, Hanting Wang, Ziyue Jiang, Long Zhou, Shujie Liu, Xize Cheng, et~al.
\newblock Wavchat: A survey of spoken dialogue models.
\newblock \emph{arXiv preprint arXiv:2411.13577}, 2024.

\bibitem[Kim et~al.(2024)Kim, Moon, Tabrizi, Lee, Mahoney, Keutzer, and Gholami]{kim2024llm}
Sehoon Kim, Suhong Moon, Ryan Tabrizi, Nicholas Lee, Michael~W Mahoney, Kurt Keutzer, and Amir Gholami.
\newblock An llm compiler for parallel function calling.
\newblock In \emph{Forty-first International Conference on Machine Learning}, 2024.

\bibitem[Kirk et~al.(2023)Kirk, Mediratta, Nalmpantis, Luketina, Hambro, Grefenstette, and Raileanu]{kirk2023understanding}
Robert Kirk, Ishita Mediratta, Christoforos Nalmpantis, Jelena Luketina, Eric Hambro, Edward Grefenstette, and Roberta Raileanu.
\newblock Understanding the effects of rlhf on llm generalisation and diversity.
\newblock \emph{arXiv preprint arXiv:2310.06452}, 2023.

\bibitem[Kong et~al.(2024)Kong, Goel, Badlani, Ping, Valle, and Catanzaro]{kong2024audio}
Zhifeng Kong, Arushi Goel, Rohan Badlani, Wei Ping, Rafael Valle, and Bryan Catanzaro.
\newblock Audio flamingo: A novel audio language model with few-shot learning and dialogue abilities.
\newblock \emph{arXiv preprint arXiv:2402.01831}, 2024.

\bibitem[Li et~al.(2024)Li, Fan, Gu, Li, Duan, Dong, Liu, and Wang]{li2024flexkbqa}
Zhenyu Li, Sunqi Fan, Yu~Gu, Xiuxing Li, Zhichao Duan, Bowen Dong, Ning Liu, and Jianyong Wang.
\newblock Flexkbqa: A flexible llm-powered framework for few-shot knowledge base question answering.
\newblock In \emph{Proceedings of the AAAI conference on artificial intelligence}, volume~38, pages 18608--18616, 2024.

\bibitem[Matena and Raffel(2022)]{matena2022merging}
Michael~S Matena and Colin~A Raffel.
\newblock Merging models with fisher-weighted averaging.
\newblock \emph{Advances in Neural Information Processing Systems}, 35:\penalty0 17703--17716, 2022.

\bibitem[Nam et~al.(2024)Nam, Macvean, Hellendoorn, Vasilescu, and Myers]{nam2024using}
Daye Nam, Andrew Macvean, Vincent Hellendoorn, Bogdan Vasilescu, and Brad Myers.
\newblock Using an llm to help with code understanding.
\newblock In \emph{Proceedings of the IEEE/ACM 46th International Conference on Software Engineering}, pages 1--13, 2024.

\bibitem[Nematullayevna(2024)]{nematullayevna2024role}
Umarova~Lobar Nematullayevna.
\newblock The role of paralinguistic cues in social life.
\newblock \emph{ANALYSIS OF MODERN SCIENCE AND INNOVATION}, 1\penalty0 (2):\penalty0 215--218, 2024.

\bibitem[Peng et~al.(2024)Peng, Wang, Xi, Li, Zhang, and Yu]{peng2024survey}
Jing Peng, Yucheng Wang, Yu~Xi, Xu~Li, Xizhuo Zhang, and Kai Yu.
\newblock A survey on speech large language models.
\newblock \emph{arXiv preprint arXiv:2410.18908}, 2024.

\bibitem[Rafailov et~al.(2023)Rafailov, Sharma, Mitchell, Manning, Ermon, and Finn]{rafailov2023direct}
Rafael Rafailov, Archit Sharma, Eric Mitchell, Christopher~D Manning, Stefano Ermon, and Chelsea Finn.
\newblock Direct preference optimization: Your language model is secretly a reward model.
\newblock \emph{Advances in Neural Information Processing Systems}, 36:\penalty0 53728--53741, 2023.

\bibitem[Rubenstein et~al.(2023)Rubenstein, Asawaroengchai, Nguyen, Bapna, Borsos, Quitry, Chen, Badawy, Han, Kharitonov, et~al.]{rubenstein2023audiopalm}
Paul~K Rubenstein, Chulayuth Asawaroengchai, Duc~Dung Nguyen, Ankur Bapna, Zal{\'a}n Borsos, F{\'e}lix de~Chaumont Quitry, Peter Chen, Dalia~El Badawy, Wei Han, Eugene Kharitonov, et~al.
\newblock Audiopalm: A large language model that can speak and listen.
\newblock \emph{arXiv preprint arXiv:2306.12925}, 2023.

\bibitem[Ruiz(2025)]{copilot}
Jeimy Ruiz.
\newblock How to debug code with github copilot.
\newblock 2025.
\newblock URL \url{https://www.anthropic.com/news/claude-3-5-sonnet}.

\bibitem[Schuller et~al.(2013)Schuller, Steidl, Batliner, Burkhardt, Devillers, M{\"u}Ller, and Narayanan]{schuller2013paralinguistics}
Bj{\"o}rn Schuller, Stefan Steidl, Anton Batliner, Felix Burkhardt, Laurence Devillers, Christian M{\"u}Ller, and Shrikanth Narayanan.
\newblock Paralinguistics in speech and language—state-of-the-art and the challenge.
\newblock \emph{Computer Speech \& Language}, 27\penalty0 (1):\penalty0 4--39, 2013.

\bibitem[Shen et~al.(2023)Shen, Song, Tan, Li, Lu, and Zhuang]{shen2023hugginggptsolvingaitasks}
Yongliang Shen, Kaitao Song, Xu~Tan, Dongsheng Li, Weiming Lu, and Yueting Zhuang.
\newblock Hugginggpt: Solving ai tasks with chatgpt and its friends in hugging face, 2023.
\newblock URL \url{https://arxiv.org/abs/2303.17580}.

\bibitem[StepFun(2025)]{stepfun_2025}
StepFun.
\newblock Stepeval-audio-360.
\newblock 2025.
\newblock URL \url{https://huggingface.co/datasets/stepfun-ai/StepEval-Audio-360}.

\bibitem[Tang et~al.(2023)Tang, Yu, Sun, Chen, Tan, Li, Lu, Ma, and Zhang]{tang2023salmonn}
Changli Tang, Wenyi Yu, Guangzhi Sun, Xianzhao Chen, Tian Tan, Wei Li, Lu~Lu, Zejun Ma, and Chao Zhang.
\newblock Salmonn: Towards generic hearing abilities for large language models.
\newblock \emph{arXiv preprint arXiv:2310.13289}, 2023.

\bibitem[Tian et~al.(2025)Tian, Zhang, Shi, Zhang, Yu, Watanabe, and Yu]{tian2025preference}
Jinchuan Tian, Chunlei Zhang, Jiatong Shi, Hao Zhang, Jianwei Yu, Shinji Watanabe, and Dong Yu.
\newblock Preference alignment improves language model-based tts.
\newblock In \emph{ICASSP 2025-2025 IEEE International Conference on Acoustics, Speech and Signal Processing (ICASSP)}, pages 1--5. IEEE, 2025.

\bibitem[Wang et~al.(2024{\natexlab{a}})Wang, Hasler, Tanneberg, Ocker, Joublin, Ceravola, Deigmoeller, and Gienger]{wang2024lami}
Chao Wang, Stephan Hasler, Daniel Tanneberg, Felix Ocker, Frank Joublin, Antonello Ceravola, Joerg Deigmoeller, and Michael Gienger.
\newblock Lami: Large language models for multi-modal human-robot interaction.
\newblock In \emph{Extended Abstracts of the CHI Conference on Human Factors in Computing Systems}, pages 1--10, 2024{\natexlab{a}}.

\bibitem[Wang et~al.(2023)Wang, Chen, Wu, Zhang, Zhou, Liu, Chen, Liu, Wang, Li, et~al.]{wang2023neural}
Chengyi Wang, Sanyuan Chen, Yu~Wu, Ziqiang Zhang, Long Zhou, Shujie Liu, Zhuo Chen, Yanqing Liu, Huaming Wang, Jinyu Li, et~al.
\newblock Neural codec language models are zero-shot text to speech synthesizers.
\newblock \emph{arXiv preprint arXiv:2301.02111}, 2023.

\bibitem[Wang et~al.(2024{\natexlab{b}})Wang, Bi, Pentyala, Ramnath, Chaudhuri, Mehrotra, Mao, Asur, et~al.]{wang2024comprehensive}
Zhichao Wang, Bin Bi, Shiva~Kumar Pentyala, Kiran Ramnath, Sougata Chaudhuri, Shubham Mehrotra, Xiang-Bo Mao, Sitaram Asur, et~al.
\newblock A comprehensive survey of llm alignment techniques: Rlhf, rlaif, ppo, dpo and more.
\newblock \emph{arXiv preprint arXiv:2407.16216}, 2024{\natexlab{b}}.

\bibitem[Wortsman et~al.(2022)Wortsman, Ilharco, Gadre, Roelofs, Gontijo-Lopes, Morcos, Namkoong, Farhadi, Carmon, Kornblith, et~al.]{wortsman2022model}
Mitchell Wortsman, Gabriel Ilharco, Samir~Ya Gadre, Rebecca Roelofs, Raphael Gontijo-Lopes, Ari~S Morcos, Hongseok Namkoong, Ali Farhadi, Yair Carmon, Simon Kornblith, et~al.
\newblock Model soups: averaging weights of multiple fine-tuned models improves accuracy without increasing inference time.
\newblock In \emph{International conference on machine learning}, pages 23965--23998. PMLR, 2022.

\bibitem[Wu et~al.(2024)Wu, Chung, Lin, Wu, Chen, Pai, Wang, Chang, Liu, and Lee]{wu2024codec}
Haibin Wu, Ho-Lam Chung, Yi-Cheng Lin, Yuan-Kuei Wu, Xuanjun Chen, Yu-Chi Pai, Hsiu-Hsuan Wang, Kai-Wei Chang, Alexander~H Liu, and Hung-yi Lee.
\newblock Codec-superb: An in-depth analysis of sound codec models.
\newblock \emph{arXiv preprint arXiv:2402.13071}, 2024.

\bibitem[Xiong et~al.(2024)Xiong, Bian, Li, Li, Du, Wang, Yin, and Helal]{xiong2024search}
Haoyi Xiong, Jiang Bian, Yuchen Li, Xuhong Li, Mengnan Du, Shuaiqiang Wang, Dawei Yin, and Sumi Helal.
\newblock When search engine services meet large language models: visions and challenges.
\newblock \emph{IEEE Transactions on Services Computing}, 2024.

\bibitem[Xu et~al.(2025)Xu, Guo, He, Hu, He, Bai, Chen, Wang, Fan, Dang, et~al.]{xu2025qwen2}
Jin Xu, Zhifang Guo, Jinzheng He, Hangrui Hu, Ting He, Shuai Bai, Keqin Chen, Jialin Wang, Yang Fan, Kai Dang, et~al.
\newblock Qwen2. 5-omni technical report.
\newblock \emph{arXiv preprint arXiv:2503.20215}, 2025.

\bibitem[Yang et~al.(2024)Yang, Tian, Tan, Huang, Liu, Guo, Chang, Shi, Bian, Zhao, et~al.]{yang2024uniaudio}
Dongchao Yang, Jinchuan Tian, Xu~Tan, Rongjie Huang, Songxiang Liu, Haohan Guo, Xuankai Chang, Jiatong Shi, Jiang Bian, Zhou Zhao, et~al.
\newblock Uniaudio: Towards universal audio generation with large language models.
\newblock In \emph{Forty-first International Conference on Machine Learning}, 2024.

\bibitem[Zeghidour et~al.(2021)Zeghidour, Luebs, Omran, Skoglund, and Tagliasacchi]{zeghidour2021soundstream}
Neil Zeghidour, Alejandro Luebs, Ahmed Omran, Jan Skoglund, and Marco Tagliasacchi.
\newblock Soundstream: An end-to-end neural audio codec.
\newblock \emph{IEEE/ACM Transactions on Audio, Speech, and Language Processing}, 30:\penalty0 495--507, 2021.

\bibitem[Zeng et~al.(2024)Zeng, Du, Liu, Wang, Jiang, Zhao, Dong, and Tang]{zeng2024glm}
Aohan Zeng, Zhengxiao Du, Mingdao Liu, Kedong Wang, Shengmin Jiang, Lei Zhao, Yuxiao Dong, and Jie Tang.
\newblock Glm-4-voice: Towards intelligent and human-like end-to-end spoken chatbot.
\newblock \emph{arXiv preprint arXiv:2412.02612}, 2024.

\bibitem[Zhang and Sennrich(2019)]{zhang2019root}
Biao Zhang and Rico Sennrich.
\newblock Root mean square layer normalization.
\newblock \emph{Advances in Neural Information Processing Systems}, 32, 2019.

\bibitem[Zhang et~al.(2023)Zhang, Li, Zhang, Zhan, Wang, Zhou, and Qiu]{zhang2023speechgpt}
Dong Zhang, Shimin Li, Xin Zhang, Jun Zhan, Pengyu Wang, Yaqian Zhou, and Xipeng Qiu.
\newblock Speechgpt: Empowering large language models with intrinsic cross-modal conversational abilities.
\newblock \emph{arXiv preprint arXiv:2305.11000}, 2023.

\bibitem[Zhang et~al.(2024)Zhang, Li, Li, Zhang, Wang, Zhou, and Qiu]{zhang2024speechalign}
Dong Zhang, Zhaowei Li, Shimin Li, Xin Zhang, Pengyu Wang, Yaqian Zhou, and Xipeng Qiu.
\newblock Speechalign: Aligning speech generation to human preferences.
\newblock \emph{arXiv preprint arXiv:2404.05600}, 2024.

\bibitem[Zhuang et~al.(2025)Zhuang, Huang, Cheng, Wu, Hu, Liao, Huang, Wang, Liao, Cai, et~al.]{zhuang2025vistorybench}
Cailin Zhuang, Ailin Huang, Wei Cheng, Jingwei Wu, Yaoqi Hu, Jiaqi Liao, Zhewei Huang, Hongyuan Wang, Xinyao Liao, Weiwei Cai, et~al.
\newblock Vistorybench: Comprehensive benchmark suite for story visualization.
\newblock \emph{arXiv preprint arXiv:2505.24862}, 2025.

\bibitem[Zhuang et~al.(2023)Zhuang, Yu, Wang, Sun, and Zhang]{zhuang2023toolqa}
Yuchen Zhuang, Yue Yu, Kuan Wang, Haotian Sun, and Chao Zhang.
\newblock Toolqa: A dataset for llm question answering with external tools.
\newblock \emph{Advances in Neural Information Processing Systems}, 36:\penalty0 50117--50143, 2023.

\bibitem[Zou et~al.(2021)Zou, Zhang, Cai, Ma, Cheng, Wang, Shi, Cheng, and Yin]{zou2021pre}
Lixin Zou, Shengqiang Zhang, Hengyi Cai, Dehong Ma, Suqi Cheng, Shuaiqiang Wang, Daiting Shi, Zhicong Cheng, and Dawei Yin.
\newblock Pre-trained language model based ranking in baidu search.
\newblock In \emph{Proceedings of the 27th ACM SIGKDD Conference on Knowledge Discovery \& Data Mining}, pages 4014--4022, 2021.

\end{thebibliography}

}

\section{Contributors and Acknowledgments}
We designate contributors as those who have been involved in the development of Step-Audio-AQAA throughout its entire process. Contributors are listed in alphabetical order by first name.
\begin{itemize}[leftmargin=1em]
    \item Research: Ailin Huang, Bingxin Li, Bruce Wang, Boyong Wu, Chao Yan, Chen Hu, Chengli Feng, Heng Wang, Hongyu Zhou, Hongyuan Wang, Jingbei Li, Jianjian Sun, Joanna Wang, Mingrui Chen, Peng Liu, Ruihang Miao, Shilei Jiang, Tian Fei, Wang You, Xi Chen, Xuerui Yang, Yechang Huang, Yuxiang Zhang, Zheng Ge, Zheng Gong, Zhewei Huang, Zixin Zhang
    \item Infra: Bin Wang, Bo Li, Buyun Ma, Changxin Miao, Changyi Wan, Chen Xu, Dapeng Shi, Dingyuan Hu, Enle Liu, Guanzhe Huang, Gulin Yan, Hanpeng Hu, Haonan Jia, Jiahao Gong, Jiaoren Wu, Jie Wu, Jie Yang, Junzhe Lin, Kaixiang Li, Lei Xia, Longlong Gu, Ming Li, Nie Hao, Ranchen Ming, Shaoliang Pang, Siqi Liu, Song Yuan, Tiancheng Cao, Wen Li, Wenqing He, Xu Zhao, Xuelin Zhang, Yanbo Yu, Yinmin Zhong, Yu Zhou, Yuanwei Liang, Yuanwei Lu, Yuxiang Yang, Zidong Yang, Zili Zhang
    \item Project Sponsors: Binxing Jiao, Daxin Jiang, Heung-Yeung Shum, Jiansheng Chen, Jing Li, Shuchang Zhou, Xiangyu Zhang, Xinhao Zhang, Yibo Zhu
    \item Corresponding: Daxin Jiang (djiang@stepfun.com), Shuchang Zhou (scotzhou@stepfun.com), Chen Hu (hatcher@stepfun.com).
\end{itemize}

\end{document}